# Event Space Theory and Its Application

## Zhang Jiangao

Faculty of Construction Management and Real Estate, Chongqing University, Chongqing 400045, China

E-mail addresses: zjiangao@yeah.net.

## Wang Weitao

Department of Computer Science, Chongqing University, Chongqing, China 400045

**Abstract**

In this paper, the basic ideal of the Event Space Theory and Analyzing Events are expatiated on. Then it is suggested that how to set up event base library in developing application software. Based above the designing principle of facing methodology. Finally, in order to explain how to apply the Event Space Theory in developing economic evaluation software, the software of "sewage treatment CAD" in a national "8th-Five Year Plan Research Project" of PRC is used as an example. This software concerns economic effectiveness evaluation for construction projects.

**1. Introduction**

There was a national 8th-Five Year Plan Research Project (sewage treatment CAD) supported by the Construction Department of the PRC in 1992,which concludes following contents: 1. building design(CAD); 2. expert systems of construction projects evaluation; 3. building budgetary estimate and budget; 4. intelligent system of economic effectiveness evaluation for construction projects. Our main work was developing the software of building budgetary estimate and budget and developing the intelligent system of economic effectiveness evaluation for construction projects.

In China, all the evaluation methods of construction projects and economy parameters should be uniform with "the methods and parameters of economic evaluation in construction projects" which is issued by State Development Planning Commission P.R.China and Ministry of Construction.

Vast time and energies will be spent for the original methods of calculation by hand and evaluation. And it is needed to recalculate economic effectiveness and redraw varies economic report tables because of the modification of data and re-estimate. At the same time, a special attention should be paid to the construction projects in different professions, new construction projects, extending construction projects and the construction projects getting financial support from foreign countries. It is because that there are different economic options, evaluation indicators and evaluation standard for above in China. In order to realize the automation and intelligence in the feasibility research of construction project gradually and make the economic effectiveness evaluation more close to the real situation of the construction project that is completed. It is suggested in this "8th-Five Year Plan Research Project" that it is needed to make the economic effectiveness evaluation construction projects by using computer software.

In order to develop the intelligent system of economic effectiveness evaluation for construction projects, we analyzed the problems of construction projects economic evaluation completely according to the economic indexes, calculating methods of economic quantities and the evaluation requires of different construction projects, which are prescribed by "the methods and parameters of economic evaluation in construction projects". We found that it seems to have definite criterions and consistency for the economic effectiveness evaluation of all construction projects. But indeed, there is quite different from project to project in evaluation methods, economic quantities calculation and calculation programs, because of the diversity and complexity of construction projects and the different requires for the economic effectiveness evaluation, and the infection of the economic data which the project could got.

The economic evaluation of a construction project is always taking place during the period of feasibility research. Therefore some of economic data can only be got from estimate and will be modified in this period. We think that the software programmed by us should be universal. It is that this software can not only do the economic evaluation for "sewage treatment" projects but also do the economic evaluation for general construction projects. However, a universal intelligent system of economic effectiveness evaluation for construction projects should have the description environment in which the users could describe their own evaluation options, methods and events. And it could be done to name them and produce menu items and bottoms. The persons that were trained professionally for economic evaluation should be familiar with the description environment.

At that period, there only are three researchers in our group, who were doing the software design and programming. WE found that it is difficult to develop a system of the economic effectiveness evaluation for universal construction projects based on the existing design theory and techniques of computer software in a short time. In order to complete this system in the period stated by the national "8th-Five Year Plan Research Project", we put forward the designing technology of facing methodology, or it is called designing technology of method base based on structure design and technology of facing objects in developing process. This technology is supported by the Event Space Theory of Software Designing and the Method of Analyzing Events. This theory and methodology might propose a basic theory and an active method for developing a software frame, which possesses a dynamic interface opened completely. It can be said that our theory is the expanding for structure design and technique of facing objects.

**2. Existing dynamic technology and methodology**

Every software programmer might know some about the dynamic technology of software design and some interface of software claimed as being a dynamic interface opened completely. But according to our opining for the opening and dynamic, there has been no software that which can achieve the complete opening and dynamic for its interface and its responsive event. In the strict sense, the opening of the interface has been only a kind of opening about interface design and a kind of predefined dynamic. This kind of predefined dynamic design could be named as secondary dynamic technique, knowing from the wide-open dynamic technology. The basic theory technique of the secondary dynamic technique is based on the technique of facing objects, whose typical design mode is the design of dynamic menus and dynamic buttons.

First, there should be a menu item or a button on the interface, whose function is to set up a new menu item or a new button. The operations or responsive events that which could be done by the new menu item or button are already predefined by the developer of the software. For example, a dynamic menu could be the below mode.

The developer has predefined some processes of treating events as follows:

```
procedure NewCommand1Click(Sender: TObject);

procedure NewCommand2Click(Sender: TObject);

      ……
```

Once a new function "NewCommandk" is defined on the interface by the user, the program codes programmed by the developer can operate the following program, set up the menu item event trigger, and put a variable MI with type TMenuItem.

```
MI := TMenuItem.Create(FileMenu);

MI.Caption := 'NewCommandk (or the name of menu item defined by user ) ';

MI.OnClick := NewCommandkClick;
```

So that the users seem to see the menu item defined by themselves and can operate the expanding function for them reserved in the software.

Generally, it is predefined in the software that the different interfaces will be showed for the different conditions and different operates. What should be done is to change the visibility of the all parts on the controlling interface.

Another method in common use is that a program environment and a set of program language similar to some high-level languages are defined by the developer. The users could program under this program environment. The problem is that these program environment and program languages might quit different from the professional usage of the application domain. Therefore, many users might be difficult to carry out the functions they want and so on.

Normally, the developer is not possible to predefine all the events the users need and to know the all the operations what users want for any application domain. So, the mode of predefinition is impossible to resolve the all problems. And predefinition will make the software too bulky and some users will feel the software too complicated. If the mode of setting up program environment and program language is used the users should possess equivalent software developing knowledge. Therefore it would be better to open the original codes rather than to define a program environment and a set of program language by the developer.

In order to develop the software which possess real dynamic opening and allow the users to expand the functions they need, what we shall do are as follows: expatiating on the basic ideal of the Event Space Theory, putting forward the concept of the Event Base, discussing the event analytical methods and how to set up the event base library of the interface event response.

In order to resolve all the problems of the application domain by using the application software, the developers only need to find the event base of the event space within the domain of the application. And they predefine the processes or functions of treating events within the base and set up a few universal abstract processes or functions for the expression of the general events under the guide of the Event Space Theory that we have expatiated. And the users could also define the functions, menu items and buttons they need. It is unthinkable for the developers of the software that what things could be treated and what functions could be possessed by the menus and buttons.

## 3. Basic Ideal of the Event Space Theory

Thinking over an application domain, any action or operation in the domain is called an event. The set of all the events in this domain is said as "event space". Every event in the event space could be decomposed into more basic events for a definite scope. In the other words, Every event in the event space is composed by the events that are more basic.

For example, an even which means to calculate the average of a field data, could be decomposed as the events as follows: selecting appointed field, ascertaining the number of the data, summing the data and dividing the sum by the number. If we suppose "calculating the average of a field data" as event A, "selecting appointed field" as event B, "ascertaining the number of the data" as event C, "summing the data" as event D and "dividing the sum by the number" as event E, A can be expressed as:

$$A = B + C + D + E,$$

where the additive order could not be commuted. This kind of additive is named as ordinal additive or non-commutative additive. Therefore the event A could be expressed as the ordinal sum of the event B, C, D, E.

If the operation is valid the average could be gained. If the operation is invalid the operation state could be could be ensured as that appointed field is not in existence or the number of the field data is zero or the addition cannot be operated for the data in the appointed field

The multiple or numeral multiple of an event could be also defined. The meanings of that an event is multiplied by a number is as follows:

If the event A could be expressed as the sum of the events $B_1$, $B_2$, $B_3$, namely as:

$$A = B_1 + B_2 + B_3, \quad \text{where } B_1 = B_2 = B_3 = B,$$

it could be explained that the event A is three times of the event B, written down as $A = 3B$. We say the event A as the result that the event B is multiplied by the number 3. In the same reason, we could define the $k$ times of the event B as that the event B is multiplied by the number $k$, it could be written as $kB$.

Analogously other operations of the events could be defined.

The event space in which the operations have been defined is denoted as $\Omega$.

**Event Base:** Let S be a set of events in the event space $\Omega$. If any event in $\Omega$ could be expressed by the events in S, and any event in S couldn't be expressed by the other events in S except by itself, then S is called an event base in the event space $\Omega$. What we mean the "expression" is not always linear. It could be nonlinear and any function relation expression.

For the event base, we call the events in this event base as basic events and the events not in this event base as non-basic events. Apparently, all non-basic events could be expressed by the basic events.

In order to investigate the event space more, we introduce the isomorphism principle first.

**Isomorphism Principle:** isomorphism means the uniformity or similarity of different systems in their structures. So that it is possible to describe quite different systems by the same mode, principles and rules. Isomorphism principle provides a gist for that the rules founded in a domain could transit into another domains reasonably and accurately.

According to the isomorphism principle, the event space should be isomorphic with some abstract space.

For a material event space, if the isomorphic abstract space and the isomorphic mapping could be found, the event base could be set up by use of known existing mathematics theory. And for every basic event in the event base, the interface response events (processes or functions) could be predefined in the software. We name all of the interface response events (processes or functions) of the software, which are corresponding to all of the basic events in the event base, as the event base library.

At one time, the treating events of abstract expressions could be predefined based on the expressions such that the common objects in the abstract space are expressed by these objects are in the base. In this way, a basic core frame of a full dynamic and opening interface could be set up. The rest software work is familiar with the most developers of software.

If an event space could not be isomorphic with an abstract space, but it could be isomorphic with a subset of the abstract space, the treating events of the event base and abstract expressions could be set up as well as. Consequently, a basic core frame of the software would be set up.

However the event space might be quite complicated and some events might be fuzzy for many application domains. Therefore it is difficult to find isomorphic mapping in some cases. In these cases we could analyze the event space directly and try to find the event base, and try to get the expression of any event use by the basic events.

It is not simple to set up the event space and set up the isomorphic mapping from the event space to an abstract space for an appointed domain. For an isomorphic abstract space it is still difficult in technique to find its base and the expression of any object expressed by the base. It is the task that the developers and professionals need to collaborate, research and try to complete each other.

According to the theory of event space, the design of software could be seen as a system which is provided with possessing feedback information, in which the professionals of many subjects cooperate each other. The following figure 1 is a simple explanation of our design idea.

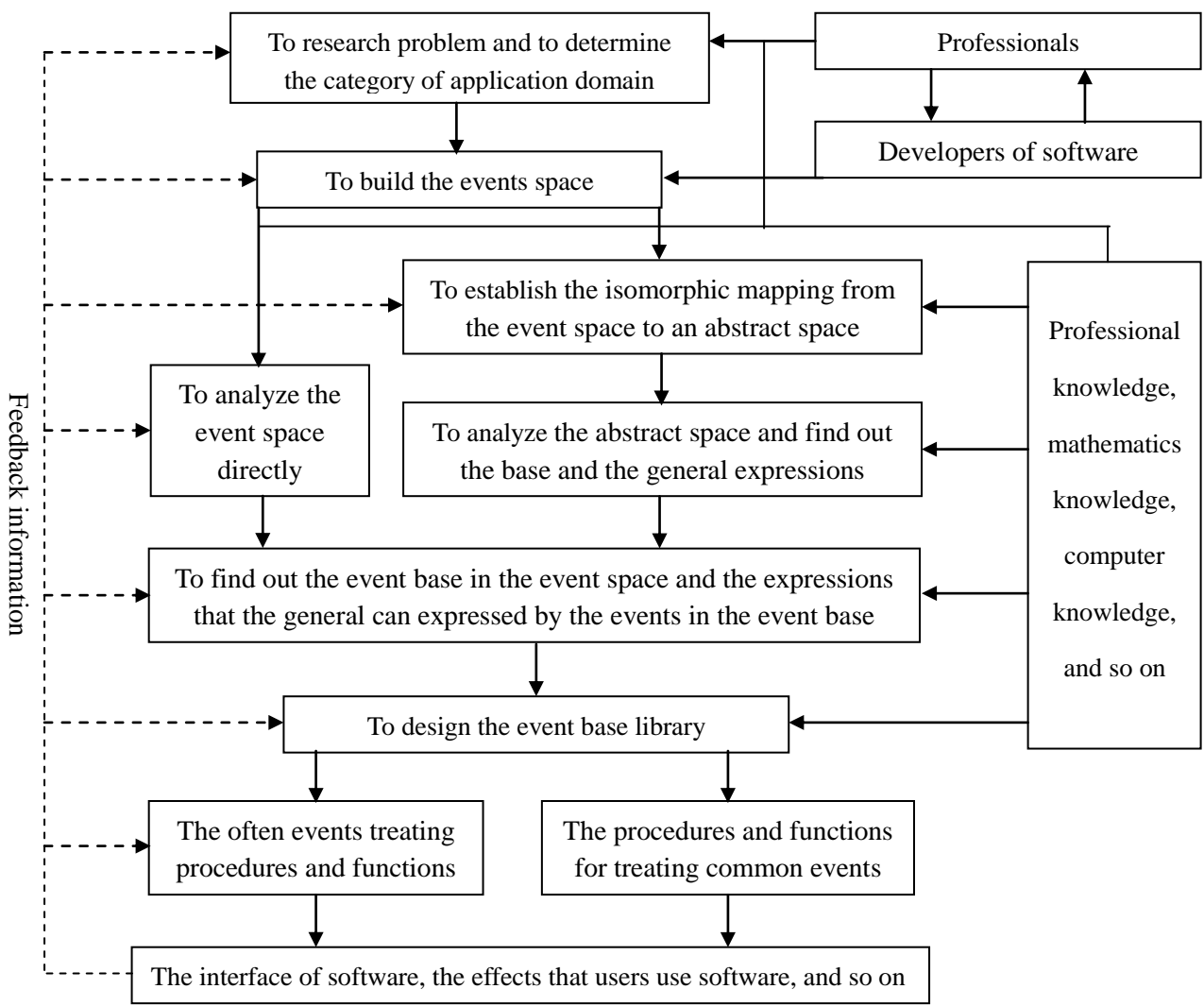

Figure 1:   A simple explanation of our design idea

## 4. More Better Expandability and Reliability

The event base library and the basic core frame of the software, which are set up by using the method of the isomorphic mapping to an abstract space, possess more better expandability and reliability compared with the methods of traditional design.

This is because that the base of an abstract space is complete. And the expressions denoted by the objects of the base for an object of the abstract space are strict and accuracy, which are guaranteed by the strict logic reasoning and proof in mathematics. Therefore the event base library based on the foundation above has the least necessary basic events and reduces the amount of code compiling in the software design, that make the re-usability of the codes reach to the highest efficiency. At the same time the basic core of the software is becoming simple and clear.

If there is material development in the application domain and the event space is expanded it is need to add some new basic events to the original event base and get the event base of the new event space. It is still simple to expand the event base library of the application software in order to adopt the new development of the application domain. What should be done is that for the adding new basic events the corresponding response events (processes or functions) of the software should be designed. In the necessary cases it might need to modify the abstract treating processes or functions of the treating event expressions, or add some treating methods for some objects.

In the traditional method, the developers might often define different codes for the events which seem non-similar, and define very complex codes for the events which seem complex because that they do not know what events are necessary. In one side it might reduce the re-usable efficiency of the codes greatly and on the other side it might reduce the operation efficiency and reliability of the software. At the same time the application software looks like an accumulation of hashes and is short of logicality and hierarchy. In this case it is not useful for debugging and maintaining, and it is not useful for the expanding too. Of cause it is impossible to achieve the full opening for event applications and interfaces because that there always are the events unimaginable by the developers. However some users might use these events quite often.

## 5. Application in the Economic Evaluation Software

We have used the design method described above in developing the software of "sewage treatment CAD" in a national "8th-Five Year Plan Research Project" of PRC. This software concerns economic effectiveness evaluation for intelligent construction projects. Some treating procedures and functions of basic events are defined directly. These procedures and functions are different each other virtually.

Because of the short time the events related to the part of displaying figures in the "uncertainty analyzing" have not been defined. So that it is impossible to translate the analyzed results and data into as figures. In this case the event treating procedures and functions defined by us have composed a non-complete event base library. But the kernel basic class of dealing with abstract event expressions is defined, in which a general dealing expression function can deal with all of the calculations for evaluation indicators and evaluation methods. Some special functions used quite often are defined, such as the function of "internal rate of return" IRR, the function of "net present value" NPV, the function of "investment period of return" IPT, the function of "investment profit ratio" IPR and so on.

The description environment and model for the "economic evaluation method of construction project" are built up. Users can describe their own evaluation options, methods and events, based on the basic events and description model proposed by us which are similar to the natural language of the professionals. And it can be done to name them and produce menu items and bottoms.

Afterwards, the basic data input frame which involve the evaluation options defined by the users will be appear on the screen when the users click the menu items or bottoms. To run these menu items and bottoms, the users can obtain the economic evaluation results of the economic items and special evaluation options pointed by users themselves.

It can be said that the software can resolve the economic evaluation subject of construction projects to a certain extent for the professionals who are similar to the application of computer for management.

# 6. Conclusion

Developing software based on the event space theory could make the inner logic structure more clear and the hierarchy more evident. At the same time the re-usability of the codes could achieve higher efficiency and increase the reliability and expandability of the software. In this case it could be easier to realize the opening of the interface and users, and the dynamic opening of functions. However, more and profounder knowledge of specialty and mathematics will be needed. Therefore the developers of the software should possess more extensive knowledge and profounder basic theory. In order to resolve a great application problem the cooperation between different subjects should be advocated.

**References**

Bertalanffy L. V., 1973. "General System Theory", George Braziller, Inc., New York.

Haken H., 1991. "Synergetic Computers and Cognition —— A Top-Down Approach to Neural Nets", Springer-Verlag Berlin Heidelberg.

Herbert Schildt, 1995. "C++: The Complete Reference, Second Edition, McGraw-Hill".

Guozhi Xu, Editor-in-Chief, 1994. The Editorial Board of Systems Science Dictionary (), "The Systems Science Dictionary", Yunnan Science and Technology Publishing.

State Development Planning Commission P.R.China and Ministry of Construction, 1995. "The Methods and Parameters of Economic Evaluation in Construction Projects", Chinese Planning Publishing.

Stefano Maruzzi, 1996. "The Microsoft Windows 95 Developer's Guide", Ziff-Davis.

Terry R. Adler, John G. Leonard, Ric K. Nordgren, 1999. Improving risk management: moving from risk elimination to risk avoidance, Information and Software Technology 41, 29-34.

Tom Swan, 1995. "Foundations of Delphi Development for Windows 95", IDG Books Worldwide, Inc..

William Roetzheim, 1994. "Programming Windows with Borland C++ 4.5", Ziff-Davis.

Yosida K., 1978. "Functional Analysis", (Chinese translated version 1981) Springer-Verlag.

Vugdelija M., Stojanović Z., Stojanovič Ž., 2000. Determination of a time step interval in hydraulic systems transients simulation, Advance in Engineering Software 31, 143-148.

Zhang X., Pham H., 2000. An analysis of factors affecting software reliability, The Journal of System and Software 50, 43-56.

**Appendix**

A very important subject was involved when we were designing the software of "sewage treatment CAD" which is a national "8th-Five Year Plan Research Project" of P.R.C.. It is the economic effectiveness evaluation for construction projects. The economic evaluation of a construction project is always taking place during the period of feasibility research. Therefore some of economic data can only be got from estimate and will be modified in this period.

Vast time and energies will be spent for the original methods of calculation by hand and evaluation. And it is needed to recalculate economic effectiveness and redraw varies economic report tables because of the modification of data and re-estimate. At the same time, a special attention should be paid to the construction projects in different professions, new construction projects, extending construction projects and the construction projects getting financial support from foreign countries. It is because that there are different economic options, evaluation indicators and evaluation standard for above in China. In order to realize the automation and intelligence in the feasibility research of construction project gradually and make the economic effectiveness evaluation more close to the real situation of the construction project that is completed. It is suggested in this "8th-Five Year Plan Research Project" that it is needed to make the economic effectiveness evaluation construction projects by using computer software.

We tried to make the software programmed by us generally. It is that this software can not only do the economic evaluation for "sewage treatment" projects but also do the economic evaluation for general construction projects. We found that it is difficult to develop a system of the economic effectiveness evaluation for general construction projects based on the existing design theory and techniques of computer software and rely on a few researchers and programmers in a sort time. Therefore we put forward the Event Space Theory of Software Designing and the Method of Analyzing Events. It proposes a basic theory and an active method for developing a software frame, which possesses a dynamic interface opened completely.

Attention is also paid to an application area. An action is called as an event. All of events in this area is named as an Event Space in which the operation of event and event base are defined. Then the isomorphism between Event Space and Abstract Space is set up based on the isomorphism principle and the isomorphism mapping is found. The dealing event of abstract expression and reacting event (procedure or function) of every basic event in the basic event base are predefined. The all of the software reacting events (process or function) which correspond with the basic events of the event base, is named as the event basic library of the software. Based on above a basic kernel frame is built up which is opened dynamically and completely.

In this paper，the Event Space Theory of Software Design and the Method of Analyzing Events are introduced and the application of this theory in the economic evaluation software for construction projects is discussed.

We have used the design methodology above to define the dealing procedures and functions of the some basic events which are different essentially each other during developing the software system of an intelligent economic effectiveness evaluation for the "sewage treatment CAD".

The kernel basic class of dealing with abstract event expressions is defined, in which a general dealing expression function can deal with all of the calculations for evaluation indicators and evaluation methods. Some special functions used quite often are defined, such as the function of "internal rate of return" IRR, the function of "net present value" NPV, the function of "investment period of return" IPT, the function of "investment profit ratio" IPR and so on.

The description environment and model for the "economic evaluation method of construction project" are built up. Users can describe their own evaluation options, methods and events, based on the basic events and description model proposed by us which are similar to the natural language of the professionals. And it can be done to name them and produce menu items and bottoms.

Afterwards, the basic data input frame which involve the evaluation options defined by the users will be appear on the screen when the users click the menu items or bottoms. To run these menu items and bottoms, the users can obtain the economic evaluation results of the economic items and special evaluation options pointed by users themselves.

It can be said that the software can resolve the economic evaluation subject of construction projects to a certain extent for the professionals who are similar to the application of computer for management.

This software system will be more reliable and more extendable than traditional design method. This is because that the base of abstract space is completed generally and the expression is strict and exact when the object of the base expresses any object, which is guaranteed by exact logical inference and proofing mathematically. Therefore, there is least needed basic events in the event base library built by the new method. This method reduces the content of code programming in software and makes the reusability reach to the highest efficiency. Meanwhile the basic kernel of the software becomes simple and clear.

In traditional design method, it does happen frequently that many events, which look not similar, are defined with different codes and the events, which look complex, are defined with complex codes. This is because that the programmers do not know which basic events are needed. The traditional design method makes the software lose logicality and hierarchy, which is not useful for debugging and extending. In the case of designing software based on the event space it is only needed to extend the event space and to add some new basic events for the original event base when the application area has been developed actually.

For example, some treating procedures (functions) of basic events and some general dealing expression function are defined directly as following:

```
const uchar
    opLBracket   = 1,    //[   (
    opRBracket   = 2,    //[   )
    opMult       = 3,    //[   *
    opDivide     = 4,    //[   /
    opSubtract   = 5,    //[   −   ......  Subtract sign
    opPlus       = 6,    //[   +   ......  Plus sign
    opPercent    = 7,    //[   %
    ……
    opAbsolute   = 101,  //[   abs(x)
    opCubeRoot   = 105,  //[   cbrt(x)
    ……
    opExp        = 109,  //[   exp(x)   ___ e^x
    ……
    opLog        = 112,  //[   log(x)
    opLog10      = 113,  //[   log10(x)
    opPower      = 116,  //[   pow(x, y) __ x^y
    opSqRoot     = 119,  //[   sqrt(x) ___ square root
    opCubic      = 122,  //[   cubic(x)
    opPoly       = 124,  //[   poly(x, c_0, c_1, ... , c_N )____The value of polynomial of degree N
                         //[   at x ; c_0, c_1, ... , c_N   are coefficients.
class Expression
{
    public:
    ……
    CalResult *   calculate( SufExpre * , ushort );   // SufExpre is a structure
    CalResult *   calculate( char *, ushort );
    CalResult *   calculate( SufExpre * );
    ……
}
```